\documentstyle[11pt]{article}
\newcommand{\bra}{\begin{array}}
\newcommand{\era}{\end{array}}
\newcommand{\beq}{\begin{equation}}
\newcommand{\eeq}{\end{equation}}
\newcommand{\bqn}{\begin{eqnarray}}
\newcommand{\eqn}{\end{eqnarray}}

\def\BC{\bb C}
\def\_\BC{\bbi C}

\def\Tr {{\rm Tr}}

\def\bz {\bar{z}}

\def\( {\left(}
\def\) {\right)}
\def\[ {\left[}
\def\] {\right]}

\def\Tr {{\rm Tr}}
\def\dag {{\dagger}}

\def\no2 {{\textstyle{n\over 2}}}

\newcommand{\om}{\omega}

\newcommand{\te}{\theta}

\newcommand{\al}{\alpha}

\newcommand{\lb}{\label}

\newcommand{\NP}[1]{ {\it Nucl.~Phys.} {\bf #1}}

\newcommand{\JMP}[1]{ {\it J. Math.~Phys.} {\bf #1}}

\begin{document}
\thispagestyle{empty}
\begin{flushright}
ucd-tpg/0602\\
hep-th/0605289
\end{flushright}

\vspace{5mm}

\begin{center}
{\Large \bf Effective Wess-Zumino-Witten Action for Edge\\
States of 
 Quantum Hall Systems on Bergman Ball  }

\vspace{5mm}

{\bf Mohammed Daoud}$^{a} $\footnote{ Permanent address :
Physics Department, Faculty of Sciences, University Ibn Zohr, Agadir,
Morocco.\\ e-mail: m$_{-}$daoud@hotmail.com} {\bf and Ahmed
  Jellal}$^b$ \footnote{ e-mail : ajellal@ictp.it -- jellal@ucd.ac.ma}\\

\vspace{0.5cm} {\em $^a$ Max Planck Institute for Physics of Complex
Systems, N\"othnitzer Str. 38, \\ D-01187 Dresden, Germany}
 \vspace{0.3cm}

{\em $^b$ Laboratory of Condensed Matter Physics, Faculty of
Sciences,  Chouaib Doukkali University,\\ P.O. Box 4056,   24000 El
Jadida, Morocco} \vspace{3cm}

{\bf Abstract}
\end{center}

Using a group theory approach, we investigate the basic features of
the Landau problem on the Bergman ball ${\bf{B}}^k$. This can be
done by considering a system of particles living on ${\bf{B}}^k$ in
the presence of an uniform magnetic field $B$ and realizing  the
ball as the coset space $SU(k,1)/U(k)$. In quantizing the theory on
${\bf{B}}^k$, we define the wavefunctions as the Wigner
$\cal{D}$-functions satisfying a set of suitable constraints. The
corresponding Hamiltonian is mapped in terms of the right
translation generators. In the lowest Landau level, we obtain the
wavefunctions as the  $SU(k,1)$ coherent states. This are used to
define the star product, density matrix and excitation potential in
higher dimensions. With these ingredients, we construct a
generalized effective Wess-Zumino-Witten action for the edge states
and discuss their nature.

\newpage

\section{Introduction and motivations}

Very recently, the quantum Hall effect (QHE)~\cite{prange} in higher dimensional spaces has
attracted much attention~\cite{zhang,group0}. The interest
on this topic is mainly motivated by its well-known significance
to the condensed matter physics~\cite{zhang}. Also its relationships to other
theories
 like non-commutative geometry and string
theory. In fact, a relation between QHE on
the complex projective spaces ${\bf CP}^k$~\cite{karabali1}
and the fuzzy spaces~\cite{knr} is established. Moreover, the string
theory is realized in terms of the QHE ingredients  in higher dimensions~\cite{fabinger}

The first results on the subject have been reported
 by Zhang and Hu~\cite{zhang} in analyzing QHE on four-sphere ${\bf S}^4$
submitted to the $SU(2)$ background field. This analysis is
motivated by the fact that the edge excitations might provide an
approach to a quantum formulation of graviton in four-dimension. QHE
in higher dimensions inherent many properties of its two-dimension
counterpart. In the conventional QHE a droplet of fermions,
occupying a certain volume, behaves as an incompressible fluid. The
low-energy excitations being area-preserving deformations and behave
as massless chiral bosons~\cite{stone,iso} described by an effective
Wess-Zumino-Witten (WZW) action in $(1+1)$-dimensions.

The derivation of the effective action for the edge states for any
quantum Hall droplet can be performed according to the method used by
Sakita~\cite{sakita2} in two-dimension Euclidean space and generalized
to ${\bf CP}^k$ by
Karabali and Nair~\cite{karabali1}. Two
ingredients are necessary to perform this derivation~: (i) a
density of states which must be constant over the phase volume occupied by
the droplet and (ii) the commutators between operators tend, in a
suitable limit (strong magnetic field), to Poisson
brackets associated to the considered manifold. More precisely, Karabali and
Nair~\cite{karabali1},
in considering QHE on ${\bf CP}^k$, gave the energy and the Landau
wavefunctions using a purely group theory approach. Furthermore, they
have shown, in an elegant way, that for a  strong magnetic field, the
quantum system is described by a constant density of states over the phase
space.
They have derived the effective action of excitations on the  ${\bf
  CP}^k$ boundaries. The obtained action is a generalized WZW action
that describes the bosonized theory of fermions in  $(1+1)$-dimensions~\cite{sakita2}.
Our motivation is based
on~\cite{karabali1} and the analytic method used by one~\cite{jellal}
of the present authors
to deal with QHE on  the Bergman ball ${\bf{B}}^k$.

We algebraically investigate
a system of particles living on the manifold  ${\bf{B}}^k$
in the presence of an uniform magnetic field
$B$. After realizing    ${\bf{B}}^k$ as the coset space $SU(k,1)/U(1)$,
we construct the wavefunctions as the Wigner $\cal{D}$-functions
verifying a set of suitable constraints. The
corresponding Hamiltonian $H$ can be written in terms of the $SU(k,1)$
generators. This will be used to define  $H$ as
a second order differential operator in the complex coordinates,
which coincides with the Maass Laplacian~\cite{elstrodt} on the
Bergman ball. To
 get the energy levels, we use the correspondence between the Landau
 problem on two manifolds   ${\bf CP}^k$ and ${\bf{B}}^k$.
We introduce an excitation potential to remove the degeneracy of the
ground state. For this, we consider a potential expressed in terms
of the $SU(k,1)$ left actions. For a strong magnetic field, we show
that the excitations of the lowest landau level (LLL) are governed
by a generalized effective Wess-Zumino-Witten (WZW) action. It turn
out that this action for $k=1$, i.e. disc, coincides with
one-chiral bosonic action for QHE at the filling factor
$\nu=1$~\cite{stone,iso}. We finally discuss the nature of the edge
excitations and in particular
 show that the field describing the edge excitations on the disc  ${\bf{B}}^1$
is
a superposition of oscillating on the boundary ${\bf{S}}^1$ of the
quantum Hall droplet.

The present paper is organized as follows.
In section 2, we present a group theory approach to analysis the Landau problem on the
Bergman ball. We build the wavefunctions and give the corresponding
Hamiltonian as well as its energy levels. In section 3, we restrict
our attention to
 LLL to write down the corresponding star product and defining the
relevant density matrix. Also we consider the excitation potential
and get the associate symbol to examine the excited states. We
determine the generalized effective
WZW action for the edge states for a strong magnetic field as well as
discuss their nature
and give the disc as example, in section 4.
We conclude and give some
perspectives in the last section.

\section{Quantization and Hilbert space}

We start by defining the Bergman ball ${\bf B}^k$, which is realized as the
coset spaces $SU(k, 1)/U(k)$. In analyzing the ${\bf B}^k$ geometry,
we derive the $U(1)$ gauge potential generating an uniform magnetic
field. In quantizing the theory, we get wavefunctions as the Wigner
${\cal D}$-functions subjected to a suitable set of
constraints. Imposing the polarization condition, we obtain the LLL
wavefunctions. The latter turn out to be the $SU(k, 1)$ coherent
states constructed from the highest weight state of the completely symmetric
representation.

\subsection{Introducing the manifold ${\bf B}^k$}

The $k$-dimensional complex ball of
unit radius is defined via $(k+1)$ homogeneous complex
coordinates $u_{\alpha}$ ($ \alpha = 1, 2, \cdots, k+1 $)
satisfying
\beq
\eta_{\alpha \beta}u^{\alpha}\bar u^{\beta} = -1
\eeq
where the $(k+1)\times(k+1)$ diagonal matrix $\eta$ is $(1,
1,\cdots ,1, -1)$. We introduce the local independent
coordinates $(z_1, z_2, \cdots , z_k)$, in the
region where $u_{k+1} \neq 0$, as $z_i = u_i/u_{k+1}$. It
follows that the global coordinates can be written in terms of
$z_i$ as
\begin{equation}
 u_\alpha = {1\over \sqrt {1-\bar z \cdot z}} \pmatrix{ z_1 \cr
z_2\cr  \vdots \cr z_k \cr 1 \cr}
\end{equation}
where  $z_i$ are verifying
\beq
\bar z \cdot z =
\sum_{i=1}^{k} \vert z_i\vert^2< 1.
\eeq
In $z$ coordinates,
the higher dimensional
complex Bergman ball ${\bf B}^k$ is defined by
\begin{equation}
{\bf B}^k = \{ z = (z_1, z_2, \cdots , z_k)\in {\bf C}^k,\ \bar z \cdot
z< 1\}.
\end{equation}
The K\"ahler potential on this manifold reads as
\begin{equation}\lb{kp}
K(\bar z, z) =- \ln( 1 - \bz  \cdot z).
\end{equation}
This generates the metric tensor
\begin{equation}
ds^2 = g_{i \bar j} dz^{i}d\bar {z}^{j}
\end{equation}
where the elements $ g_{i \bar j}$ are related to $K(\bar z, z) $ by
\begin{equation}
 g_{i \bar j} = \frac{\partial^2 K(\bar z, z)}{\partial z^{i}\partial \bar
 {z}^{j}}.
\end{equation}
Using (\ref{kp}), we explicitly write  $ g_{i \bar j}$ as
\begin{equation}
 g_{i \bar j} = \frac{\delta_{ij}}{1 - \bar z \cdot z} + \frac{ \bar z_i \cdot z_j}{(1 - \bar
 z \cdot z)^2}
\end{equation}
This implies
\begin{equation}\lb{metric}
ds^2 =  \frac{dz d\bar z}{(1 - \bar z \cdot z)^2} + \frac{(z \cdot d\bar
z)(\bar z \cdot dz)}{(1 - \bar z \cdot z)^4}.
\end{equation}
The ball ${\bf B}^k$ is equipped with a closed two-form
 $\omega$, such as
\begin{equation}\lb{tf}
\omega = i g_{i \bar j} dz^i \wedge d\bar z^j.
\end{equation}
 Since  $\omega$ is non-degenerate, it is a symplectic two-form
and therefore allows us to construct a Poisson bracket.
This is
\begin{equation}\lb{pb}
\{ f_1 , f_2 \} = i g^{\bar i j}\left( \frac{\partial f_1 }{\partial z_j
} \frac{\partial f_2 }{\partial \bar z_i } -  \frac{\partial f_1
}{\partial \bar z_i }\frac{\partial f_2 }{\partial z_j }\right)
\end{equation}
where $g^{\bar i j}$  is the inverse of tensor metric
\begin{equation}
g^{\bar i j} =  (1 - \bar z.z)(\delta_{ij} - \bar z_i \cdot z_j)
\end{equation}
In terms of the local coordinates, (\ref{pb}) can be written as
\begin{equation}
\{ f_1 , f_2 \} = -i(1 - \bar z \cdot z)
 \left( \frac{\partial f_1 }{\partial z_i }
\frac{\partial f_2 }{\partial \bar z_i } -  \frac{\partial f_1
}{\partial \bar z_i }\frac{\partial f_2 }{\partial z_i } -z
\frac{\partial f_1 }{\partial z }\bar z \frac{\partial f_2 }{\partial
\bar z } + \bar z  \frac{\partial f_1 }{\partial \bar
z}z \frac{\partial f_2 }{\partial z } \right).
\end{equation}
To make contact with the group theory, we note that
the manifold ${{\bf B}}^k$ can be viewed as the coset space $SU(k,1)/U(k)$.
This provides us with the algebraic tools to deal with QHE on  the
Bergman ball ${{\bf B}}^k$.

Let be $g$ the $(k+1)\times(k+1)$  matrices of a fundamental representation of the group
$SU(k,1)$, with $g \in SU(k,1)$. They obey the relations
\begin{equation}
\eta g^+ \eta = g^{-1},\qquad \det g = 1.
\end{equation}
Considering the $t_{\alpha}$  generators of $SU(k,1)$ as matrices in
the fundamental representation, such as
\beq
2\Tr \left(t_{\alpha}t_{\beta}\right) = \delta_{\alpha \beta}
\eeq
and the $t_i$ generators of $SU(k)\subset U(k)$ as matrices having zero for
the $(k+1)^{\sf th}$ row and column, with $i= 1, 2, \cdots , k^2-1$. The generator corresponding to the
$U(1)$ direction of the subgroup $U(k)$ will be denoted by
$t_{k^2+2k}$. It can be written as
\begin{equation}
t_{k^2+2k}  = \frac{1}{\sqrt{2k(k+1)}}\pmatrix{{\bf{I}}_k & 0\cr 0&
-k\cr}
\end{equation}
where ${\bf I}_k$ is the $(k\times k)$ unit matrix.
Note that, an element $g$ of $SU(k,1)$ can be used to
parametrize ${\bf B}^k$ by identifying $g$ to $gh$, with $h \in
U(k)$.

In terms of $g$, the one-form, i.e. U(1)
connection, is given by
\begin{equation}\lb{theta}
\theta = i \sqrt {\frac{2k}{k+1}}\ \Tr\left(t_{k^2+2k} g^{-1}dg\right).
\end{equation}
Using the result
\beq
\Tr\left(g^{-1}dg\right)=0
\eeq
and setting $u_{\alpha} = g_{\alpha,k+1}$,
we show that
\begin{equation}
\theta = i \eta _{\alpha \beta} \bar u^{\alpha}du^{\beta}.
\end{equation}
It can be written in the local coordinates as
\begin{equation}\lb{teta}
\theta = \frac{i}{2}\frac{\bar z  \cdot dz  - z  \cdot d\bar z}{1 - \bar z \cdot z}.
\end{equation}
This formula agrees with (\ref{tf}) because it is easy to see
that $\omega = d\theta$. Note that,~(\ref{teta}) will play an
important role in quantizing the theory describing a system
living on the manifold ${\bf B}^k$. Indeed, as we are interested
to analysis QHE on the higher dimensional
ball, it is necessary to identify the magnetic field behind
this phenomena. This field is proportional to the
K\"ahler two-form~(\ref{tf}) and since  $\om$ is closed,
the components of the magnetic field expressed in terms of the frame
fields defined by the metric are constants.

\subsection{Quantization}

A basis of functions on $SU(k,1)$ is given by the Wigner $\cal{D}$-functions
\begin{equation}\lb{wf}
{\cal D}_{L,R}^K (g)= \langle K , L | g |K , R \rangle.
\end{equation}
They are the matrix elements corresponding to $g$ in
the discrete representation of $SU(k,1)$. The quantum numbers
$L$ and $R$ are specifying the states on which the $SU(k,1)$ generators
act. The left and right $SU(k,1)$ actions are defined by
\begin{equation}\lb{lraction}
L_{\alpha} g = t_{\alpha} g, \qquad R_{\alpha} g = g
t_{\alpha}.
\end{equation}
These relations will play a crucial role in defining two different
Hamiltonians, see next.

The Hilbert space corresponding to the quantum system living on the
manifold ${\bf B}^k$ can be determined by reducing the degrees of
freedom of the system on $SU(k,1)$ to the coset space
$SU(k,1)/U(k)$. This reduction can be realized by
imposing a set of suitable constraints. 
%
%
%
To find the constraints arising in the quantization of the theory,
let us consider
the $U(1)$ gauge field, i.e. the vector potential
associated to the field strength $F$, as
\begin{equation}\lb{pvect}
A = - i n \eta_{\alpha \beta } \ \bar u_{\alpha} \ du_{\beta} 
\end{equation}
where $n$ is a positive real quantum number. Hereafter, we assume that
$n$ is an integer since we are considering the discrete series of the
unitary representation corresponding to the group $ SU(k,1)$.
The gauge can be also written as
\begin{equation}
A = - i n  \sqrt {\frac{2k}{k+1}}\ \Tr\left(t_{k^2+2k} g^{-1}dg\right).
\end{equation}
Note that, under the gauge transformation $g
\longrightarrow gh$ where $h \in SU(k)$, we have $A
\longrightarrow A$. However,
Under the $U(1)$ transformation
\beq
g \longrightarrow g\exp\left(it_{k^2+2k}\varphi\right)
\eeq
we have
\beq
A \longrightarrow A
+d\left(n \sqrt {\frac{k}{2(k+1)}}\varphi\right).
\eeq
It follows that under the
gauge transformation $g \longrightarrow gh$ where $h \in U(k)$,
the wavefunctions~(\ref{wf}) transform as
\begin{equation}
{\cal D}_{L,R}^K (gh) = \exp\left(\int dt \dot{A}\right) {\cal D}_{L,R}^K (g)
\end{equation}
where $\dot{A}$ is given by
\begin{equation}
\dot{A} = - i n  \sqrt {\frac{2k}{k+1}}\ \Tr\left(t_{k^2+2k}
h^{-1}\dot{h}\right).
\end{equation}
Thus, one can see that the canonical momentum corresponding
to the $(k^2 + 2k)$-direction is quantized as $n k/\sqrt{2k(k+1)}$. Consequently,
the admissible quantum states generating the Hilbert space must satisfy the constraint
\begin{equation}\lb{aqs1}
R_{k^2 + 2k} {\cal D}_{L,R}^K (g) = \frac{nk}{\sqrt{2k(k+1)}}{\cal
D}_{L,R}^K (g).
\end{equation}
Furthermore, due to the invariance of $U(1)$ gauge field under
$SU(k)$ transformations, the corresponding canonical momentum are
vanishing. This invariance leads to the set of constraints
\begin{equation}\lb{aqs2}
R_{j} {\cal D}_{L,R}^K (g) = 0, \qquad j = 0,
1, \cdots , k^2-1.
\end{equation}
We denote by
$t_{-i} $ and $ t_{+i}$ the  $SU(k,1)$ generators, which do not belong
to the group $U(k)$,
with $i = 1, 2, \cdots, k$. They
can be seen as
the lowering and raising operators analogously to the annihilation and
creation operators of the harmonic oscillator.
They will be next related to the covariant
derivatives on ${\bf B}^k$.
From  (\ref{aqs1}-\ref{aqs2}) and using the commutation
relations of the algebra of $SU(k,1)$, we obtain
\begin{equation}\lb{aqs3}
\left[R_{-i}, R_{+j}\right] = n \delta_{ij}.
\end{equation}
The constraints (\ref{aqs1}-\ref{aqs2})
mean that the wavefunctions are singlets under the right $SU(k)$
action and carry a right $U(1)$ charge induced by the background
field. Because  of this, in the decomposition of $SU(k+1)$ irreducible
representations into $SU(k)$ ones, we should consider the
representations satisfying~(\ref{aqs1}-\ref{aqs2}).

It is important to
note that,~(\ref{aqs1}-\ref{aqs3}) are similar to those
obtained
by Karabali and Nair~\cite{karabali1} by considering the quantization of the
complex projective spaces ${\bf CP}^k = SU(k+1)/U(k)$.
Consequently, we apply the representation theory analysis developed
in~\cite{karabali1}
to  the ball ${\bf B}^k$ case. This can be done by using the standard
procedure associating a compact group with
his non-compact analogue. We will return to this matter in
deriving the spectrum  of the system.

\subsection{Lowest Landau levels}

To find the
lowest Landau levels, we should require the condition
\begin{equation}\lb{lllc}
R_{-i} {\cal D}_{L,R}^K (g) = 0
\end{equation}
which is known as the polarization condition in the context of the
geometric quantization. Therefore,  the wavefunctions of the
LLL belong to the symmetric representations of
$SU(k,1)$.

The complete symmetric $SU(k,1)$ representations can be realized
in terms of $(k+1)$-bosons satisfying the commutation relations
\begin{equation}
\left[ a_r , a_s^{\dag}\right] = \delta_{rs}, \qquad  r,s = 0,1, \cdots , k.
\end{equation}
The restricted Fock  space, i.e. representation space, is defined by
\begin{equation}\lb{fock}
{\cal F} = \left\{ \|K, n_1, n_2, \cdots,
 n_k\rangle \equiv  |n_0, n_1,
n_2, \cdots, n_k \rangle, \  n_0 - (n_1+ n_2+ \cdots+ n_k)= 2K-1 \right\}
\end{equation}
where the
states  are given by
\begin{equation}\lb{nstates}
|n_0, n_1, n_2 \cdots, n_k \rangle =
\frac{(a_0^{\dag})^{n_0}(a_1^{\dag})^{n_1}(a_2^{\dag})^{n_2}\cdots (a_k^{\dag})^{n_k}
}{\sqrt{n_0!n_1!n_2!\cdots n_k!}} \
 |K, 0, 0, 0, \cdots, 0 \rangle
\end{equation}
and the vacuum is annihilated by the operator $a_r$
\begin{equation}
a_r|K, 0, 0, 0, \cdots, 0 \rangle = 0.
\end{equation}
We can realize
the lowering $t_{-i}$ and raising $t_{+i}$
operators as
\begin{equation}\lb{tcr}
t_{-i} = a_0 a_i, \qquad t_{+i} = a_0^{\dag} a_i^{\dag}, \qquad
1 \leq i \leq k.
\end{equation}
They satisfy the relation
\begin{equation}\lb{tcom}
\left[t_{-i} , t_{+j}\right] = a_0 a_0^{\dag}\delta_{ij} +  a_j^{\dag} a_i.
\end{equation}
The Lowest Landau condition~(\ref{lllc}) implies that the right quantum
numbers are $n_i = 0$ for $1 \leq i \leq k$ and $n_0 =
K$. Acting~(\ref{tcom}) on the states
$|n_0, 0, 0, \cdots, 0 \rangle$, we get $ 2K$. Now using the
commutation relation~(\ref{aqs3}), we obtain the relation $K={n\over 2}$.
Hereafter, we set
\beq
B= 2n
\eeq
where $B$ stands for the strength of the magnetic field.

The coset generators act on the Fock space, associated to the
symmetric representation, as
\bqn
\lb{tact}
t_{-i}\|K, n_1, n_2, \cdots,n_i,\cdots
 n_k\rangle =\sqrt{n_i\left(2K-1+(n_1+ n_2+\cdots\+n_k)\right)}\ \|K, n_1, n_2, \cdots,n_i-1,\cdots
 n_k\rangle, \nonumber\\
t_{+i}\|K, n_1, n_2, \cdots,n_i,\cdots
 n_k\rangle =\sqrt{(n_i+1)\left(2K+(n_1+ n_2+\cdots\+n_k)\right)} \ \|K, n_1, n_2, \cdots,n_i+1,\cdots
 n_k\rangle
\eqn
 From the previous analysis, the
wavefunctions corresponding to LLL are given by
\begin{equation}\lb{lllf}
\psi_{LLL} = \langle K, n_1,n_2,\cdots , n_k \| g \| K, 0, 0 , \cdots,
0 \rangle.
\end{equation}
As far as  the manifold ${\bf B}^k$ is concerned, we identify $g$
in~(\ref{lllf}) with the unitary exponential mapping $\Omega$ of the space
generated by lowering and raising operators
\begin{equation}\lb{Omega}
\sum_{i=1}^k \left(\eta_i t_{+i} - \bar \eta_i t_{-i}\right)\longrightarrow
\Omega = \exp\left\{\sum_{i=1}^k \left(\eta_i t_{+i} - \bar \eta_i t_{-i}\right)\right\}
\end{equation}
where $\eta_i$, $i= 1, 2, \cdots, k$ are complex parameters and
$\Omega$ is an unitary coset
representative of  the ball ${\bf B}^k \equiv SU(k,1)/U(k)$, i.e.
\beq
\Omega^{\dag}\Omega = 1.
\eeq
Using~(\ref{tact}),
we show that the action of $\Omega$ on the lowest weight vector $\| K,
0, 0 , \cdots , 0
\rangle$ leads
\begin{equation}\lb{lllwf}
\psi_{LLL} =
\sqrt{\frac{(n-1+n_1+\cdots +n_k)!}{(n-1)!\ n_1!n_2!\cdots
n_k!}} \ \frac{z_1^{n_1}z_2^{n_2}\cdots z_k^{n_k}}{(1-\bar z \cdot
z)^{-\frac{n}{2}}}
\end{equation}
where the complex variables $z_i$ are related to the
parameters $\eta_i$ in (\ref{Omega}) by
\beq
z_i = \eta_i \tanh\left (\sqrt{\bar \eta \cdot
\eta} \right)/\sqrt{\bar \eta  \cdot \eta}
\eeq
and  the $(1\times k)$-matrix $\bar
\eta $ is
\beq
\bar \eta =  \left [\bar  \eta_1  \cdot \bar \eta_2   \cdots \ \bar
\eta_k \right].
\eeq
It follows that LLL are
infinitely degenerated. The wavefunctions
$\psi_{LLL}\equiv \psi_{n_1, \cdots ,n_k}(z_1, \cdots , z_k )$
obey the orthogonality condition
\begin{equation}
\int d\mu(z_1, \cdots , z_k) \
 \bar \psi_{n'_1,  \cdots ,n'_k}(z_1, \cdots , z_k ) \ \psi_{n_1, \cdots ,n_k}(z_1, \cdots ,
z_k )= \delta_{n'_1,n_1}\cdots \delta_{n'_k,n_k}
\end{equation}
where the measure is given by
\begin{equation}\lb{mu}
 d\mu(z_1, \cdots , z_k) =
 (n-k)  (n-k+1) \cdots  (n-1) \ {d^2z_1 \cdots
 d^2z_k\over \pi^{k} (1-\bar z\cdot z)^{k+1}}
\end{equation}
Note that,
the representation space $\cal{F}$ has an infinite
dimension. At this point, one may ask for the energy levels
corresponding to the
wavefunctions constrained by~(\ref{aqs1}-\ref{aqs3}).
The answer can
be given  by defining the relevant  Hamiltonian that describes
the quantum system living on the ball ${\bf B}^k$.

\section{Hamiltonian and spectrum}

After quantizing the theory, we have obtained the wavefunctions of the
quantum system on  ${\bf B}^k$. The corresponding Hamiltonian can be defined
by making use of the structure relation (\ref{aqs3}) arising from the
constraints~(\ref{aqs1}-\ref{aqs2}). To illustrate the present analysis, we
consider the disc  ${\bf B}^1$, i.e. $k=1$, as a particular case.

\subsection{Hamiltonian }

We begin by writing, in the
parametrization $(z_i, \bar{z}_i)$, the Hamiltonian describing a
non-relativistic particle living on the manifold ${\bf B}^k$
in  the presence of the $U(1)$ magnetic field given by (19). To obtain the
discrete spectrum, we use a group
theoretical approach based on the results presented before. It is
important
to note that, the $2k$ right generators $R_{\pm i}$
of $SU(k,1)$ can be related to the covariant
derivatives on ${\bf B}^k$.
The right actions behave like
creation and annihilation operators for the standard harmonic
oscillators. Thus, it is natural to define the corresponding
Hamiltonian as
\begin{equation}\lb{rham}
H =  \frac{1}{2} \sum_{i=1}^{k} \left(R_{+i}R_{-i}+ R_{-i}R_{+i}\right).
\end{equation}
It is useful to identify  the elements $g \in
SU(k,1)$ to $gh$ ($h \in U(k)$) in order to view their corresponding
two points as one point in ${\bf B}^k$. Thus, the Maurer-Cartan one-form $g^{-1}dg$
can be only expressed in terms of the coset coordinates
($z_i$, $\bar z_i$). It is given by
\begin{equation}\lb{cmof}
g^{-1}dg = - i t_{+i}\ e^i_j \ dz^j - i t_{-i}\ e^{\bar i}_{\bar j}\ d\bar
z^j - i t_{a}\ e^a_j \ dz^j - i t_{a} \ e^{a}_{\bar j} \ d\bar z^j
-2i\sqrt{\frac{k+1}{2k}}t_{k^2+2k}\ \theta.
\end{equation}
The $U(1)$ connection $\theta$~(\ref{theta}) can be
 rewritten as
\beq
\theta = \theta_i dz^i + \theta_{\bar i}d\bar
z^{\bar j}
\eeq
where the components $\te_i$ and $\theta_{\bar i}$ are
\beq
\theta_i = {i \bar z_i \over 2(1-\bar z  \cdot z) }, \qquad
\theta_{\bar i}= -{i z_i \over 2(1-\bar z \cdot z)}.
\eeq
In one-form (\ref{cmof}),
we have introduced the complex vielbeins
\beq
 e^i = e^i_j dz^j, \qquad e^{\bar i} = e^{\bar i}_{\bar j}d\bar z^{j}.
\eeq
They are defined such that  the line element (\ref{metric}) takes the form
\begin{equation}
ds^2 = e^i \delta_{i \bar i}e^{\bar i}
\end{equation}
and related to the metric by
\begin{equation}
g_{i \bar j} = e^k_i \delta_{k \bar k}e^{\bar k}_{\bar j}.
\end{equation}
The explicit expressions of the vielbeins components are
\bqn
\lb{viel}
e^k_j &=& \frac{i}{\sqrt{1-\bar z  \cdot z}} \ \left( \delta_{kj} -
\frac{z_k \cdot
\bar z_j }{\bar z \cdot z}\right) + \frac{i}{1-\bar z  \cdot
  z}\frac{z_k\cdot \bar
z_j}{\bar z \cdot z},\nonumber\\
e^{\bar k}_{\bar j} &=& -\frac{i}{\sqrt{1-\bar z \cdot z}}\ \left(
\delta_{kj} - \frac{z_j \cdot \bar z_k }{\bar z \cdot z}\right) -\frac{i}{1-\bar
z \cdot z}\frac{z_j \cdot \bar z_k}{\bar z \cdot z}.
\eqn
In obtaining this result, we have used a similar approach to that elaborated in~\cite{hirch}
for ${\bf CP}^k$. One can
verify that the metric inverse is
\begin{equation}
g^{i \bar j} = (e^{-1})^i_k \delta_{k \bar k}(e^{-1})^{\bar
j}_{\bar k}
\end{equation}
which will be used in deriving the analytical realization of the
Hamiltonian.
 From (\ref{cmof}), we obtain
\bqn
\lb{e-1}
\left(e^{-1}\right)^j_i\frac{\partial g}{\partial z^j} &=& g \left[ -it_{+i} -
it_a e^a_j(e^{-1})^j_i
-2i\sqrt{\frac{k+1}{2k}}t_{k^2+2k}\theta_j(e^{-1})^j_i
\right],\nonumber \\
\left(e^{-1}\right)^{\bar j}_{\bar i}\frac{\partial g}{\partial \bar z^j} &=& g
\left[ -it_{-i} - it_a e^a_{\bar j}(e^{-1})^{\bar j}_{\bar i}
-2i\sqrt{\frac{k+1}{2k}}t_{k^2+2k}\theta_{\bar j}(e^{-1})^{\bar
j}_{\bar i} \right].
\eqn
One can see that the $SU(k,1)$ right actions, defined by (\ref{lraction}),
can be written as
\begin{equation}\lb{rreal}
R_{+i} = i \left(e^{-1}\right)^j_i D_j, \qquad R_{-i} = i
\left(e^{-1}\right)^{\bar j}_{\bar i} D_{\bar j}
\end{equation}
where the $U(1)$ covariant derivatives $D_j$ and $D_{\bar j}$ are
given by
\begin{equation}\lb{CD}
D_j = \frac{\partial }{\partial z^j} - i A_j, \qquad D_{\bar
j} = \frac{\partial }{\partial \bar z^j} - i A_{\bar j}.
\end{equation}
The quantities
\beq
A_j = - i\frac{n}{2}\bar z_j (1-\bar z \cdot z)^{-1}, \qquad
A_{\bar j} =  i\frac{n}{2} z_j (1-\bar z \cdot z)^{-1}
\eeq
are the
components of the potential vector, in $z$ coordinates,
\beq
A = A_jdz_j +
A_{\bar j}dz_{\bar j}
\eeq
 given by~(\ref{pvect}). To obtain the formula (\ref{CD}), we
have used the constraints~(\ref{aqs1}-\ref{aqs2}) those must be satisfied by
the physical wavefunctions of the present system. Putting (\ref{rreal}) in (\ref{rham}), we
find
\begin{equation}\lb{dham}
H = -\frac{1}{2}\left(g^{i \bar j} D_i D_{\bar j} + g^{\bar i j}
D_{\bar i}D_j\right)
\end{equation}
where we have a summation over the repeated indices.
In the complex coordinates, we straightforwardly show that the Hamiltonian~(\ref{dham})
 takes the form
\begin{equation}\lb{zham}
H = -(1-\bar z.z)\bigg\{(\delta_{ij} - z_i\cdot \bar z_j)\frac{\partial
}{\partial z^j}\frac{\partial }{\partial \bar
z^j}+\frac{n}{2}(z_i\frac{\partial }{\partial z^i}-\bar
z_j\frac{\partial }{\partial \bar z^j})\bigg\} + \frac{n^2}{4}
\bar z.z
\end{equation}
This coincides with the Maass Laplacian~\cite{elstrodt} in higher
dimensional spaces. It was investigated at many occasions, using the spectral
theory~\cite{group1}, in order to describe the quantum system in the presence of a
magnetic background field
$n$. Also it was analytically considered in analyzing QHE on the
Bergman ball
 ${\bf B}^k$~\cite{jellal}.

\subsection{Spectrum}

The derivation of the energy eigenvalues of the Hamiltonian
(\ref{zham}) is based on the correspondence between the Landau problem on
two manifolds ${\bf CP}^k$ and ${\bf B}^k$.
Indeed, as we claimed before, the constraints~(\ref{aqs1}-\ref{aqs2})  together
with the relation~(\ref{aqs3}) are similar to those obtained in analyzing
the Landau system on  ${\bf CP}^k$~\cite{karabali1}.
Thus, to get the
discrete spectrum of a non-relativistic charged particle on ${\bf
B}^k$, we use the standard procedure to pass from ${\bf B}^k$ to
${\bf CP}^k$. For this, we set
\begin{equation}
R_{+i} = i J_{+i}, \qquad R_{-i} = i J_{-i}, \qquad
R_{a} =  J_{a}, \qquad R_{k^2+2k}
= J_{k^2+2k}
\end{equation}
where $a$ runs over the set $\{1,\cdots,k^2-1\}$.
The $J$'s satisfy the structure relations characterizing the algebra
of $SU(k+1)$. Now,
 the relations~(\ref{aqs1}-\ref{aqs3}) become
\begin{equation}
 J_{a} \psi = 0, \qquad
  J_{k^2+2k} \psi = \frac{nk}{\sqrt{2k(k+1)}}\psi, \qquad
 [J_{+j} , J_{-i}] \psi = n\delta_{ij}\psi
\end{equation}
which are nothing but the relations involved in the Landau systems
on ${\bf CP}^k$ where the $U(1)$ generator corresponding to
$(k^2+2k)$-direction takes the fixed value $n k /\sqrt{2k(k+1)}$. Then,
according to~\cite{karabali1}, the representations of $SU(k+1)$,
which contain $SU(k)$ singlets, are labeled by two integers $(p,q)$
where the condition
\beq
q-p=n
\eeq
 must fulfilled. The second order Casimir of $SU(k+1)$
\beq
C_2 =
\sum_{\al=1}^{k^2+2k} J_{\al}^2
\eeq
is given by
\begin{equation}
C_2 = \frac{k}{2(k+1)}\left[p(p+k+1) + q(q+k+1) +
\frac{2}{k}pq\right].
\end{equation}
Writing the Hamiltonian $H$ (\ref{rham}) as
\begin{equation}
H =  -\frac{1}{2} \sum_{i=1}^{k} (J_{+i}J_{-i}+ J_{+i}J_{-i})
\end{equation}
one can easily show  that
\begin{equation}
H =  J_{k^2+2k}^2 - C_2.
\end{equation}
Finally, we find the eigenvalues of $H$ as
\begin{equation}\lb{evalue}
E_q = \frac{n}{2}(2q+k)-q(q+k)
\end{equation}
which agrees with the analytical analysis~\cite{jellal}.
The index $q$ labels the Landau levels.
LLL
has the energy
\beq
E_0 = {nk\over 2}
\eeq
 and the corresponding wavefunctions are
given by~(\ref{lllwf}).
This value coincides with that can be obtained by considering the Landau
problem
on the real space ${\bf R}^{2k}$.
We close this section by giving an example.

\subsection{Example : disc $\bf{B}^1$}

To illustrate the strategy adopted in obtaining
the energy spectrum~(\ref{evalue}), we consider the simplest case
$k=1$~\cite{daoud1}, i.e. the disc ${\bf B}^{1}$.
Firstly, we derive the spectrum using the $SU(1,1)$ discrete
representations and secondly, we compare the obtained result with that
 derived in passing from the
non-compact group $SU(1,1)$ to its compact partner $SU(2)$.

In the Cartan-Weyl basis, the  Lie algebra of
of $SU(1,1)$ is characterized by the commutation relations
\begin{equation}\lb{trc1}
 \left[ t_3 , t_{\pm}\right] = \pm t_{\pm}, \qquad \left[t_{-} , t_{+}\right] =
 2t_3.
\end{equation}
The positive discrete series representations of $SU(1,1)$ are
labeled by an integer $l$. The representation space is generated  by
the states $\{| l, q\rangle\}$. They are
eigenstates of $t_3$ with  $l+q$ as eigenvalues. The
constraint~(\ref{aqs1}) reduces  to
\beq
R_3 \psi = \frac{n}{2} \psi
\eeq
and implies that
\beq\lb{dicc}
l+q=
\frac{n}{2}.
\eeq
The eigenvalues of the Hamiltonian
\beq\lb{diham}
H =
\frac{1}{2} \left(R_+R_- + R_-R_+\right)
\eeq
 in the above representation are
\beq
E=
\frac{n^2}{4} - l(l-1)
\eeq
where $l(l-1)$ is the eigenvalue of the
$SU(1,1)$ second order Casimir. From the constraint~(\ref{dicc}), we get
\begin{equation}\lb{evalue1}
E_q = \frac{n}{2}(2q+1)-q(q+1)
\end{equation}
which agrees  with~(\ref{evalue}) for $k=1$.

We now consider the
passage from $SU(1,1)$ to $SU(2)$. For this, we set
\beq
t_3 = j_3, \qquad j_{\pm} = i t_{\pm}.
\eeq
The relations (\ref{trc1}) become
those of the algebra of $SU(2)$, such as
\begin{equation}
 [ j_3 , j_{\pm}] = \pm j_{\pm}, \qquad [j_{+} , j_{-}] =
 2j_3.
\end{equation}
This mapping allows us to write the Hamiltonian
(\ref{diham}) as
\beq\lb{hamc}
H = j_3^2 - C_2\left[SU(2)\right] = {n^2\over 4} - j(j-1)
\eeq
where ${n\over 2}$ and $ j(j-1)$ are the eigenvalues of  $j_3$ and
$C_2$, respectively. For a given
$SU(2)$ unitary irreducible representation $j$, we have
\beq
j= q- {n\over
  2}.
\eeq
Replacing in (\ref{hamc}), we get the energies
(\ref{evalue1}). This shows the equivalence between the first and
second ways in deriving the spectrum of the system living on the disc ${\bf B}^{1}$.

\section{Lowest Landau level analysis}

Restricting to LLL, described by the wavefunctions $\psi_{LLL}$ given
in (\ref{lllwf}) of fundamental energy $E_0={nk\over 2}$, we develop
some tools needed in studying the edge excitations of the quantum Hall
droplets on ${\bf B}^k$. These concern
the star product, density of states and excitation potential.

\subsection{The star product}

For a strong magnetic field $(B\sim n)$, the particles are
confined in LLL (\ref{lllwf}). In this limit,
we will discuss how the non-commutative geometry occurs and
 show that the commutators between two operators acting on the
states $\psi_{LLL}$ give a Poisson brackets of type (\ref{pb}). To do this,
we
associate the function
\begin{equation}
{\cal A}(\bar z, z) = \langle z | A | z \rangle = \langle 0
|\Omega^{\dag} A  \Omega| 0 \rangle
\end{equation}
to any operator $A$ acting on the LLL wavefunctions.
$\Omega$ is given by (\ref{Omega}), $| 0 \rangle = \parallel K, 0, 0,
\cdots, 0\rangle$ is the lowest highest weight state of the
complete symmetric representation of $SU(k,1)$ defined in (\ref{fock}) and
the vector states
\beq
| z \rangle =\Omega| 0 \rangle \equiv | z_1,
z_2, \cdots, z_k \rangle
\eeq
 are the $SU(k,1)$ coherent states
\begin{equation}\lb{css}
| z \rangle  = (1-\bar z \cdot  z)^{\frac{n}{2}}
\sum_{n_1=0}^{\infty}
\cdots
\sum_{n_k=0}^{\infty}
\sqrt{\frac{(n-1+n_1+\cdots+n_k)!}{(n-1)!\ n_1!n_2!\cdots
n_k!}}z_1^{n_1}z_2^{n_2}\cdots z_k^{n_k}\parallel n_1, n_2,
\cdots, n_k\rangle
\end{equation}
Next, to simplify our notations, we set $\parallel n_1,
n_2, \cdots, n_k\rangle \equiv |\{ n \}\rangle$.

An associative
star product of two functions ${\cal A}(\bar z, z)$ and ${\cal
B}(\bar z, z)$ is defined by
\begin{equation}\lb{sp1}
{\cal A}(\bar z, z)\star {\cal B}(\bar z, z) = \langle z | AB | z
\rangle.
\end{equation}
Using the unitarity of $\Omega $  and
the completeness relation
\beq
\sum_{\{n\}=0}^{\infty}|\{ n
\}\rangle \langle \{ n \}|= {\bf I}
\eeq
we write (\ref{sp1}) as
\begin{equation}\lb{sp2}
{\cal A}(\bar z, z)\star {\cal B}(\bar z, z) =
\sum_{\{n\}=0}^{\infty} \langle 0 |\Omega^{\dag} A \Omega |\{ n
\}\rangle \langle \{ n \}|\Omega^{\dag} B \Omega| 0 \rangle.
\end{equation}
From~(\ref{tact}), for large $n $, the star
product becomes
\begin{equation}\lb{sp3}
{\cal A}(\bar z, z)\star {\cal B}(\bar z, z) = {\cal A}(\bar z, z)
{\cal B}(\bar z, z) + \frac{1}{n}\sum_{i=0}^{k} \langle 0
|\Omega^{\dag} A \Omega t_{+i} |0\rangle \langle
0|t_{-i}\Omega^{\dag} B \Omega| 0 \rangle + O\left(\frac{1}{n^2}\right).
\end{equation}
It is clear that,
the first term in r.h.s. of~(\ref{sp3}) is the
ordinary product of two functions ${\cal A}$ and ${\cal B}$. However, the
non-commutativity is encoded in  the second term.

To completely determine  the
star product (\ref{sp3}) for large $n$,  we should evaluate the matrix
elements of type
\beq
 \langle 0 |\Omega^{\dag} A \Omega t_{+i}
|0\rangle.
\eeq
Using the holmorphicity condition
\beq
R_{-i}\langle \{ n
\}|\Omega |0\rangle = 0
\eeq
 which can be easily verified by using the coherent states (\ref{css}), we have
\begin{equation}
 \langle 0 |\Omega^{\dag} A \Omega t_{+i}
|0\rangle = R_{+i} \langle 0 |\Omega^{\dag} A \Omega  |0\rangle.
\end{equation}
Similarly, we obtain
\begin{equation}
 \langle
0|t_{-i}\Omega^{\dag} B \Omega| 0 \rangle = - R_{-i}\langle
0|\Omega^{\dag} B \Omega| 0 \rangle
\end{equation}
where we have considered the relation
\beq
R_{+i}^{\star} = - R_{-i}.
\eeq
From
(\ref{e-1}-\ref{rreal}), since we are concerned with a  $U(1)$ abelian gauge
field, we show that (\ref{sp3}) becomes
\begin{equation}\lb{sp4}
{\cal A}(\bar z, z)\star {\cal B}(\bar z, z) = {\cal A}(\bar z, z)
{\cal B}(\bar z, z) - \frac{1}{n}g^{j \bar m}
\partial_{j}{\cal A}(\bar z, z)\partial_{\bar m}{\cal B}(\bar z,
z) + O\left(\frac{1}{n^2}\right).
\end{equation}
Therefore, the symbol or function associated to the commutator of two
operators $A$ and $B$ can be written as
\begin{equation}\lb{sp5}
\langle z |[ A , B] | z \rangle = - \frac{1}{n}g^{j \bar m}
\{\partial_{j}{\cal A}(\bar z, z)\partial_{\bar m}{\cal B}(\bar z,
z) - \partial_{j}{\cal B}(\bar z, z)\partial_{\bar m}{\cal A}(\bar
z, z)\}.
\end{equation}
This implies
\begin{equation}\lb{sp6}
\langle z |[ A , B] | z \rangle =  \frac{i}{n} \left\{{\cal A}(\bar z,
z), {\cal B}(\bar z, z)\right\} \equiv \left\{{\cal A}(\bar z, z), {\cal
B}(\bar z, z)\right\}_{\star}
\end{equation}
where $\{ , \}$ stands for the Poisson bracket (\ref{pb}) defined on the Kahlerian
manifold ${\bf B}^k$ and the Moyal bracket $\{ ,
\}_{\star}$ is given by
\beq
\{{\cal A}(\bar z, z), {\cal B}(\bar z,
z)\}_{\star} = {\cal A}(\bar z, z)\star {\cal B}(\bar z, z) -
{\cal B}(\bar z, z)\star {\cal A}(\bar z, z).
\eeq
The obtained star product shows that how the non-commutative geometry occurs in
analyzing
QHE on the Bergman ball  ${\bf B}^k$.

\subsection{Density matrix}

Another important physical quantity, needed in deriving the WZW
action for the edge excitations, is the density matrix. Since LLL are infinitely
degenerated, one may fill the LLL states with
\beq
M = M_1+M_2+\cdots +M_k
\eeq
particles where $M_i$ stands for the particle number in the mode
$i$. The corresponding density operator is then
\begin{equation}
\rho_0 = \sum_{\{m\}}| \{m\} \rangle \langle \{m\} |.
\end{equation}
Its associated symbol reads as
\begin{equation}
\rho_0(\bar z, z) = (1-\bar z \cdot z)^n \sum_{m_1 = 0}^{M_1}\cdots
\sum_{m_k = 0}^{M_k} \frac{(n-1+m_1\cdots+m_k)!}{(n-1)!\ m_1!\cdots
m_k!}|z_1|^{2m_1}\cdots |z_k|^{2m_k}.
\end{equation}
By using the identity
\begin{equation}\lb{iden}
(1-\bar z \cdot z)^{-n} =  \sum_{m_1 = 0}^{\infty}\cdots \sum_{m_k =
0}^{\infty} \frac{(n-1+m_1\cdots+m_k)!}{(n-1)\ !m_1!\cdots
m_k!}|z_1|^{2m_1}\cdots |z_k|^{2m_k}
\end{equation}
and for a strong magnetic field $(B\sim n)$,
 $\rho_0(\bar z, z) $
 can be approximated by
%
\begin{equation}\lb{density}
\rho_0(\bar z, z) \simeq \exp(-n\bar z \cdot z)\sum_{m=0}^M \frac{(n\bar
z \cdot z)^m}{m!}\simeq \Theta (M - n\bar z \cdot z)
\end{equation}
where we have set
\beq
m=m_1+\cdots+m_k.
\eeq
The obtained expression of $\rho_0(\bar z, z) $ is valid for a large number $M$ of
particles~\cite{sakita1} as well.
The mean value of the density operator is a step
function for $n\longrightarrow \infty$ and $M\longrightarrow
\infty$ ($\frac{M}{n}$ fixed). It corresponds to an abelian
droplet configuration with a boundary defined by
\beq\lb{boun}
n\bar z \cdot z = M
\eeq
and
its radius is proportional to $\sqrt{M}$. The derivative of this
density tends to a $\delta$-function. This result will be useful
in describing the edge excitations.

\subsection{Excitation potential}

The quantum Hall droplet under consideration is specified by the
density of states $\rho_0(\bar z, z) $. The excitations of this configuration can
be described by an unitary time evolution operator $U$. It gives
information concerning the dynamics of the excitations around
$\rho_0$. The excited states will be characterized by a density
operator
\beq
\rho = U \rho_0 U^{\dagger}.
\eeq
In this situation, one can
write the Hamiltonian as
\begin{equation}\lb{pham}
{\cal H} = E_0 + V
\end{equation}
where $E_0 = \frac{kn}{2}$ is the LLL energy and $V$ is the
excitation potential. The perturbation  $V$ will induces  a lifting
of the LLL degeneracy. Note that, the $SU(k,1)$ left actions commute
with the covariant derivatives. They correspond to the magnetic
translations on ${\bf B}^k$ and behind the degeneracy of the Landau
levels. Thus, it is natural to construct $V$ in terms of
the magnetic translations.

In LLL, the
$SU(k,1)$ left actions act on the complete symmetric
representations
and admit a bosonic realization, similar to that given by
(\ref{tcr}-\ref{tcom}) for the right actions.
A simple
choice for the excitation potential $V$
is
\begin{equation}
V = \omega \sum_{i=1}^{k} a_i^{\dag}a_i.
\end{equation}
One can verify
\begin{equation}
\langle K, n_1,\cdots,n_k\|V\|K, n_1,\cdots,n_k\rangle = \omega
(n_1+\cdots+n_k)
\end{equation}
which is reflecting the degeneracy lifting of the ground state.
The symbol associated to the perturbation $V$ is given by
\begin{equation}\lb{fpot}
 {\cal V}(\bar z, z ) = \langle z |V| z \rangle = n\omega \frac{\bar z
 \cdot z}{1-\bar z \cdot z}
\end{equation}
where we have used the definition (\ref{nstates}) and a relation of type
(\ref{iden}). This essentially goes to the harmonic oscillator
potential for large $n$ on the real $2k$-dimensional spaces.

\section{Edge excitations and generalized WZW action}

We now derive the effective Wess-Zumino-Witten action for the edge
states. The derivation is based on the LLL analysis given in the
previous section. As mentioned above, the dynamical information,
related to degrees of freedom of the edge states, is contained in
the unitary operator $U$.

\subsection{WZW Action}
The action, describing the edge excitations  in the
Hartree-Fock approximation, can be writing as \cite{sakita2}
\begin{equation}\lb{saction}
S = \int dt\ \Tr \left\{ \rho_0 U^{\dag}\left(i\partial_t -{\cal H}\right)U \right\}
\end{equation}
where ${\cal H}$ is given by~(\ref{pham}). For a strong magnetic
field, large $n$, the quantities occurring in (\ref{saction}) can be
evaluated as classical functions. To do this, we adopt a
method similar to that used by Karabali and Nair~\cite{karabali1}. This is mainly based on the strategy
used by Sakita~\cite{sakita2} in dealing with a bosonized theory of
fermions.

 We start by computing the term $ i \int dt\ \Tr\left(\rho_0
U^{\dag}\partial_tU\right)$. For this, we set
\beq
U= e^{+i\Phi}, \qquad \Phi^{\dag} = \Phi.
\eeq
By a direct calculation, 
we can write
\begin{equation}
U^{\dag}dU = i \int_0^1 d\tau e^{-i\tau\Phi}d\Phi
e^{+i\tau\Phi}
\end{equation}
which leads
\begin{equation}\lb{expre}
 i \int dt \ \Tr\left(\rho_0
U^{\dag}\partial_tU\right) = \sum_{n=0}^{\infty} \frac{-
(i)^n}{(n+1)!}\ \Tr
\left(\underbrace{[\Phi,\cdots[\Phi}_n,\rho_0]\cdots]\partial_t\Phi \right).
\end{equation}
Due to the completeness  of the LLL levels, the trace of any
operator $A$ is
\beq
\Tr A = \int d\mu\ \langle z | A | z \rangle
\eeq
where the measure $d\mu$ is given by equation (\ref{mu}). It becomes
clear that (\ref{expre}) can be also written in the following
form
\begin{equation}
 i \int dt\ \Tr(\rho_0
U^{\dag}\partial_tU) =\int d\mu \sum_{n=0}^{\infty} \frac{-
(i)^n}{(n+1)!}
\underbrace{\{\phi,\cdots\{\phi}_n,\rho_0\}_{\star}\cdots\}_{\star}\star\partial_t\phi.
\end{equation}
This is more suggestive for our purpose. Indeed, using the relations
(\ref{sp4}-\ref{sp6}), it is easy to see that
\begin{equation}
 i \int dt\ \Tr\left(\rho_0
U^{\dag}\partial_tU\right) \simeq \frac {1}{2n} \int d\mu\
\{\phi,\rho_0\}\partial_t\phi
\end{equation}
where we have dropped the terms in $\frac{1}{n^2}$ as well as that related to
the total time derivative. Note that, the symbol $\{ , \}$ is the
Poisson bracket (\ref{pb}) and gives
\begin{equation}\lb{vpb}
\{\phi , \rho_0\} = ({\cal L}\phi) \  \frac{\partial\rho_0(\bz, z)}{\partial
(\bar z \cdot z)}
\end{equation}
where the first order differential operator $ {\cal L}$ is
\begin{equation}
 {\cal L} =  i (1 - \bar z \cdot z)^2 \left(z \cdot \frac{\partial}{\partial z}
- \bar z \cdot \frac{\partial}{\partial \bar z}\right).
\end{equation}
Recall that, for large $n$, since the density (\ref{density}) is a step
function, its derivative is a $\delta$-function with a support on
the boundary $\partial {\cal D}$ of the droplet ${\cal D}$ defined
by~(\ref{boun}). Then, we obtain
\begin{equation}\lb{1term}
 i \int dt \ \Tr\left(\rho_0
U^{\dag}\partial_tU\right) \approx -\frac{1}{2} \int_{\partial {\cal
D}\times{\bf R}^+} dt\ ({\cal L}\phi)\ ( \partial_t\phi ).
\end{equation}

The second step in the derivation of the edge states action consists
in simplifying the term involving the Hamiltonian ${\cal H}$.
By a straightforward calculation, we obtain
\begin{equation}\lb{vterm}
\Tr\left(\rho_0 U^{\dag} V U\right) = \Tr\left(\rho_0  V \right) + i \Tr\left([\rho_0, V] \Phi\right)
+ \frac{1}{2}\Tr\left([\rho_0, \Phi ][V, \Phi ]\right).
\end{equation}
The first term in r.h.s of (\ref{vterm}) is $\Phi$-independent. We drop
it since does not contains any information about the dynamics of the
edge excitations. In terms of the Moyal bracket, the second term  in
r.h.s of (\ref{vterm})
rewrite as
\begin{equation}
 i \Tr\left([\rho_0, V] \Phi\right)  \approx i \int d\mu\ \{\rho_0,
 {\cal V}\}_{\star} \phi.
\end{equation}
Using  (\ref{fpot}), one show that
\begin{equation}
 i \Tr\left([\rho_0, V] \Phi\right)\longrightarrow 0
\end{equation}
The last term  in r.h.s of (\ref{vterm})  can be evaluated in a
similar way to
get (\ref{vpb}). As result, we have
\begin{equation}
  \int dt \ \Tr\left(\rho_0
U^{\dag}{\cal H}U\right) = - \frac{1}{2n^2}\int d\mu \ ({\cal L}\Phi)
\frac{\partial\rho_0}{\partial (\bar z \cdot z)} \ ({\cal L}\Phi)
\frac{\partial {\cal V}}{\partial (\bar z \cdot z)}
\end{equation}
Note that, we have eliminated the term containing the ground state
energy $E_0$, because does not contribute to the edge dynamics. For large
$n$, we have
\beq
\frac{\partial {\cal V}}{\partial (\bar z \cdot z)}
\longrightarrow n\omega
\eeq
and using the spatial shape of
density $\rho_0(\bz, z)$, to obtain
\begin{equation}\lb{2term}
 \int dt\ \Tr\left(\rho_0
U^{\dag}{\cal H}U\right) = \frac{\omega}{2}\int_{\partial {\cal D}\times
{\bf R }^+} dt\ ({\cal L}\phi)^2.
\end{equation}
Combining (\ref{1term}) and (\ref{2term}), we end up with the
appropriate action
\begin{equation}\lb{faction}
S \approx -\frac{1}{2}\int_{\partial {\cal D}\times {\bf R }^+} dt \
\left\{({\cal L}\phi)( \partial_t\phi )+\omega ({\cal
L}\phi)^2\right\}.
\end{equation}
This action involves only the time derivative of $\phi$ and the
tangential derivative ${\cal L}\phi$. It
is similar to that derived by Karabali and Nair~\cite{karabali1} on
${\bf CP}^k$. (\ref{faction}) shows that how
a generalized
chiral abelian WZW theory
can be constructed on the Bergman ball ${\bf B}^k$. For $k=1$, we
recover the WZW action for the edge states of hyperbolic quantum
Hall systems~\cite{daoud1}.

\subsection{Nature of excitations}

It will be interesting to
investigate the nature of edge states.
Indeed, from~(\ref{faction}), the equation of motion of the field
$\phi$ is given by
\begin{equation}\lb{geom}
{\cal L}\left( \partial_t\phi + \omega
{\cal L}\phi\right)= 0.
\end{equation}
Solving this equation, one can obtain all information concerning the
edge excitation.
More precisely, let us consider the simplest case, i.e. the disc
${\bf B}^1 = \{z\in {\bf C}, \
|z|<1\}$.

For $k=1$,  (\ref{geom})
simplifies as
 \begin{equation}\lb{eom}
\partial_{\al}\left( \partial_t\phi + \omega
\partial_{\al}\phi\right)= 0
\end{equation}
where $\al$ such that $z=|z| e^{i\al}$.
The eigenstates of the angular momenta ${\cal L} = \partial_{\al}$ take
the forms
\beq
\bar z^qz^p = |z|^{p+q}e^{i(p-q)\al}
\eeq
and the corresponding
eigenvalues are  $i(p-q)$. Note that, the zero-momentum states. i.e.
$p=q$, does not give a deformation of the boundary defined
by (\ref{boun}). The field $\phi$ can be expanded in terms of
the eigenstates of the differential operator ${\cal L}$ as
\begin{equation}
\phi (\al,t ) = \sum_{p\neq q}
a_{pq}(t)\ e^{i(p-q)\al}.
\end{equation}
Injecting this form of $\phi$ in (\ref{eom}), we show that  the time-dependent functions $a_{pq}(t)$
satisfy the following differential  equation
\begin{equation}
\partial_t a_{pq}(t) + i(p-q)\omega a_{pq}(t)=0
\end{equation}
which can be easily solved. It follows that, the edge field $\phi$
is
\begin{equation}
 \phi (\al, t ) = \sum_{p\neq q}
a_{pq}e^{i(p-q)\omega t}e^{i(p-q)\al}
\end{equation}
where the coefficients $a_{pq}$ are time-independent. This shows that the field
$\phi$ is a superposition of oscillating modes on the boundary
${\bf{S}}^1$ of the quantum Hall droplet.

\section{Conclusion}

By considering a system of particles living on the Bergman ball
 ${\bf B}^k$ in the presence of a $U(1)$ background field, we have
 algebraically investigated QHE. This was based on the fact that
 ${\bf B}^k$  can be viewed as the coset space  $SU(k,1)/
  U(1)$. This was used to get wavefunctions as the Wigner ${\cal D}$-functions
submitted to a set of suitable constraints. Also to map the
 corresponding Hamiltonian in terms of the  $SU(k,1)$ right
 generators. This latter is showed to coincide with the generalized Maass
 Lapalacian in the complex coordinates. The Landau levels on  ${\bf B}^k$
 are obtained by using the correspondence of two manifolds
${\bf CP}^k$ and
${\bf B}^k$. More precisely, we have used a mapping between the
 $SU(k,1)$ and   $SU(k+1)$ generators. In
the lowest Landau levels (LLL), the obtained wavefunctions were
 nothing but
  the   $SU(k,1)$ coherent states.

Restricting to LLL, we have derived a generalized effective WZW action that
describes the quantum Hall droplet of radius proportional to
$\sqrt{M}$, with $M$ is the number of particles in LLL. In order to
obtain the boundary excitation action, we have defined the
star product and the density of states. Also we have introduced the
perturbation potential responsible of the degeneracy lifting in terms
of the magnetic translations of $SU(k,1)$.
Finally, we have discussed the nature of the edge excitations
and illustrated this discussion by giving the disc as example.

The present analysis gives an idea how to deal with QHE on higher
dimensional non-compact spaces, i.e. the Bergman ball ${\bf B}^k$. It will be interesting to apply the
obtained results in order to deal with other issues~\cite{daoud2}.

\section*{Acknowledgment}

MD work's was partially done  during a visit to
the Max Planck Institute for Physics of
Complex Systems, Dresden--Germany. He would like to acknowledge the
financial support of the Institute.
The authors are indebted to the
referee for his constructive comment.

\end{document}